\documentclass[aps,prd,nofootinbib,twocolumn,superscriptaddress,preprintnumbers,balancelastpage,longbibliography]{revtex4-2}
\usepackage{aas_macros}

\usepackage[colorlinks=true,linkcolor=blue,citecolor=blue,urlcolor=blue]{hyperref}
\usepackage{orcidlink}
\usepackage{placeins}
\usepackage{lipsum}

\usepackage{graphicx}
\usepackage[normalem]{ulem}
\usepackage{amssymb}

\usepackage{graphicx}
\usepackage{dcolumn}
\usepackage{bm}
\usepackage{xcolor}

\def\lsim{\mathrel{\raise.3ex\hbox{$<$\kern-.75em\lower1ex\hbox{$\sim$}}}}
\def\gsim{\mathrel{\raise.3ex\hbox{$>$\kern-.75em\lower1ex\hbox{$\sim$}}}}

\newcommand{\dd}{\mathrm{d}}

\begin{document}

\makeatletter
\typeout{Column width: \the\columnwidth}
\typeout{Text width: \the\textwidth}
\makeatother

\title{Constraining The Neutrino-Nucleon Cross Section with the Ultrahigh-Energy KM3NeT Event}

\author{Toni Bertólez-Martínez\orcidlink{0000-0002-4586-6508}}
    \email{bertolezmart@wisc.edu}

\author{Dan Hooper\orcidlink{0000-0001-8837-4127}}
    \email{dwhooper@wisc.edu}
\affiliation{
    Department of Physics, Wisconsin IceCube Particle Astrophysics Center, University of Wisconsin, Madison, Wisconsin 53706, USA
}

\begin{abstract}

KM3NeT's detection of a muon track produced by a $\sim 220 \, {\rm PeV}$ neutrino provides an opportunity to probe physics at center-of-momentum energies greater than those probed by the Large Hadron Collider or other existing particle accelerators. In this study, we use this single event to place an upper limit on the neutrino-nucleon cross section of $\sigma_{\nu N} < 26 \, \sigma_{\nu N}^{\rm SM}$ at $E_{\rm CM} \sim 20 \, {\rm TeV}$. This result can be used to constrain a variety of scenarios beyond the Standard Model. With future very large volume neutrino telescopes, constraints on the neutrino-nucleon scattering cross section are expected to become substantially more stringent and, in some scenarios, could become competitive with accelerator probes of new physics.

\end{abstract}

\maketitle


\section{Introduction}
Recently, the KM3NeT Collaboration reported the detection of a muon track with an energy of $120^{+110}_{-60} \, {\rm PeV}$~\cite{KM3NeT:2025npi}. Given the exceptional energy and approximately horizontal direction of this track, it is very unlikely to be of atmospheric origin and has instead been interpreted as the product of an astrophysical neutrino. This event represents the highest energy neutrino detected to date (for further discussion, see Refs.~\cite{KM3NeT:2025aps,KM3NeT:2025ccp,Li:2025tqf,Fang:2025nzg}).

The energy of the observed muon can be used to estimate the energy of the parent neutrino. From simulations, the median energy of the neutrino responsible for this event has been estimated to be 220 PeV, with a corresponding 68\% confidence interval of 110 to 790 PeV~\cite{KM3NeT:2025npi}. This corresponds to an energy of $E_{\rm CM} \approx \sqrt{2 m_p E_{\nu}} \approx 14.4-38.5 \, {\rm TeV}$ in the center-of-momentum frame of the neutrino-nucleon collision, placing it beyond the range of energies studied at the Large Hadron Collider and other accelerator experiments.

Neutrinos arriving from near the horizon traverse more matter than down-going neutrinos, increasing the probability of their interaction. This factor is especially important for ultra-high energy neutrinos, as the resulting muons can travel as far as tens of kilometers before falling below the energy threshold of large volume neutrino telescopes. In addition, the Earth becomes opaque to neutrinos at extremely high energies, leading to the suppression of up-going neutrino-induced events. This causes the angular distribution of neutrino-induced events to depend on the total neutrino-nucleon cross section at a given energy, and allows high-energy neutrino telescopes to be sensitive to any new physics that might contribute to this cross section~\cite{Kusenko:2001gj,Hooper:2002yq,Anchordoqui:2005pn,Borriello:2007cs,IceCube:2020rnc,IceCube:2017roe,Bustamante:2017xuy,Klein:2019nbu,Denton:2020jft,Huang:2021mki,Valera:2022ylt,Zhou:2019frk}. 

\begin{figure}
    \centering
    \includegraphics[width=1\linewidth]{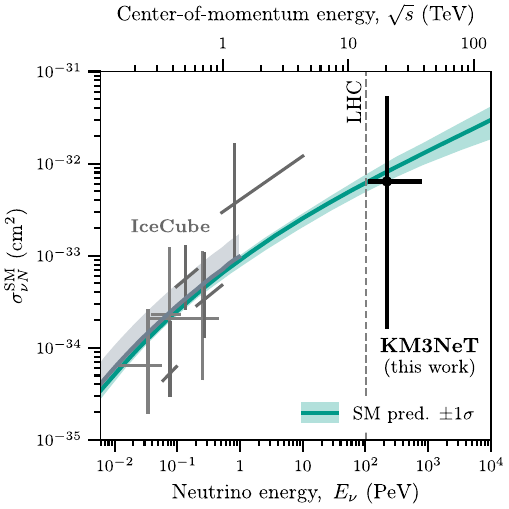}
    \caption{Measurements of the total high-energy neutrino-nucleus cross section compared to the Standard Model prediction~\cite{Bertone:2018dse}. In gray, we show the measurements from IceCube showers~\cite{Bustamante:2017xuy,IceCube:2020rnc} and tracks~\cite{IceCube:2017roe}. For measurements of the charged-current cross section, we have rescaled the results by $(\sigma^{\rm CC}_{\nu N}+\sigma^{\rm NC}_{\nu N})/\sigma^{\rm CC}_{\nu N}$, as predicted by the Standard Model. The vertical dashed line denotes the center-of-momentum energy of proton-proton collisions at the Large Hadron Collider. In black, we show the measurement of the total neutrino cross section as derived in this analysis from the KM3NeT event. \emph{For the first time, neutrino telescopes allow for a constraint of this cross section beyond the energy range of particle accelerators.}}
    \label{fig:cross-section}
\end{figure}

In this paper, we consider the ultrahigh-energy event reported by the KM3NeT Collaboration in this context. From this single event, we find that we can place significant constraints on the total neutrino-nucleon cross section at $E_{\rm CM} \sim 20 \, {\rm TeV}$ (see Fig.~\ref{fig:cross-section}). In particular, the fact that this muon track is nearly horizontal can be used to place an upper limit of $\sigma_{\nu N} < 26 \,\sigma_{\nu N}^{\rm SM}$ at the 95\% confidence level (for analogous measurements at somewhat lower energies derived from IceCube data, see Refs.~\cite{IceCube:2017roe,Bustamante:2017xuy,IceCube:2020rnc}). This information, in turn, can be used to place constraints on a variety of exotic physics scenarios, including those featuring large or warped extra dimensions~\cite{Jain:2000pu,Alvarez-Muniz:2001efi,Alvarez-Muniz:2002snq,Jain:2002kz,Anchordoqui:2002vb,Anchordoqui:2003jr,Ringwald:2001vk,Illana:2004qc}, leptoquarks~\cite{Anchordoqui:2006wc,Dey:2015eaa,Dutta:2015dka,Becirevic:2018uab,Chauhan:2017ndd,Mileo:2016zeo,Dey:2016sht,Kirk:2023fin,Romero:2009vu,Bai:2025pef}, leptogluons~\cite{Bai:2025pef}, string excitations~\cite{Cornet:2001gy,Friess:2002cc}, supersymmetric resonances~\cite{Carena:1998gd}, new gauge bosons~\cite{Pandey:2019apj}, or non-perturbative electroweak interactions~\cite{Han:2003ru,Ellis:2016dgb,Morris:1993wg,Fodor:2003bn} (for an early review, see Ref.~\cite{Han:2004kq}).

\section{The Angular Distribution of Ultrahigh-Energy Neutrino Events}

In the deep-inelastic scattering regime, the Standard Model neutrino-nucleon cross section is approximately given by~\cite{Gandhi:1995tf}
\begin{equation}
   \sigma^{\rm SM}_{\nu N} \equiv \sigma^{\mathrm{CC}}_{\nu N}+\sigma^{\mathrm{NC}}_{\nu N} \approx 8.76\times 10^{-33}\text{ cm}^2 \times \left(\frac{E_\nu}{220\text{ PeV}}\right)^{0.402}.  \\ 
\end{equation}
The corresponding cross section for antineutrinos exhibits a similar energy dependence and is only slightly smaller in magnitude. Due to the growth of this cross section with energy, ultrahigh-energy neutrinos are efficiently absorbed as they propagate through the Earth, with a mean free path that is given by
\begin{equation}\label{eq:uhe-nu-mfp}
    \lambda = \frac{1}{n_{N}\sigma_{\nu N}^{\rm SM}} \approx 673 \, {\rm km} \times \left(\frac{\rho_{\rm crust}}{\rho}\right)\left(\frac{220\, \mathrm{PeV}}{E_\nu}\right)^{0.402},
\end{equation}
where $\rho$ is the local matter density and $\rho_{\rm crust} = 2.84\,  \mathrm{g}/\mathrm{cm}^{3}$ is the mean mass density of the Earth's crust. This scattering suppresses the flux of up-going ultrahigh-energy neutrinos in telescopes such as KM3NeT and IceCube. At $E_{\nu} \approx 220 \, {\rm PeV}$, neutrinos traverse more than one mean free path of Earth for all incoming angles more than $3^\circ$ below the horizon.

Muons produced through charged-current neutrino interactions can travel a significant distance before losing the bulk of their energy. In the energy range of interest, muons traveling through water lose energy at an average rate of $\langle \dd E_{\mu}/\dd x \rangle= (0.3 \, {\rm PeV/km}) \times (E_{\mu}/{\rm PeV})$~\cite{Dutta:2000hh}. Further taking into account the stochasticity of the energy losses in this regime, we find that an EeV muon, for example, will typically travel through $\sim 6-12\, {\rm km}$ of water before its energy falls below 10 PeV~\cite{koehne2013proposal}. Since this significantly exceeds KM3NeT's depth of 3.4 km, a large fraction of ultrahigh-energy events are expected to originate from directions near the horizon.

\begin{figure}
    \centering
    \includegraphics[width=1\linewidth]{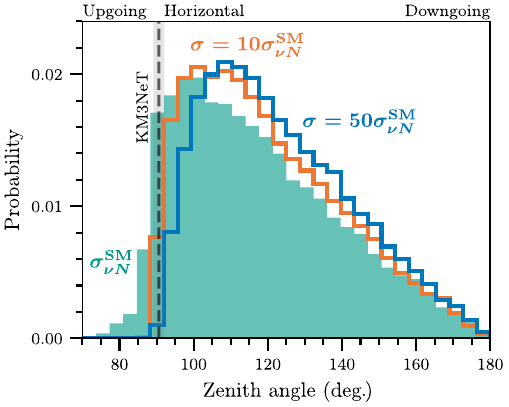}
    \caption{The angular distribution of $>$10 PeV muon tracks predicted at KM3NeT from the charged-current interactions of an isotropic flux of $\sim 220 \, {\rm PeV}$ neutrinos. Results are shown for the Standard Model cross section (filled green), and for a total cross section that is larger by a factor of 10 (orange) or 50 (blue). The vertical dashed line denotes the zenith angle of the 220 PeV KM3NeT event, $\theta_{\rm obs} = 90.6^\circ$, which is only slightly above the horizon. 
    \emph{The horizontal direction of this track is compatible with the angular distribution predicted for the Standard Model cross section, but such events would have been more likely to be downgoing if the cross section were larger.}}
    \label{fig:angular-distribution}
\end{figure}

To determine the predicted angular distribution of ultrahigh-energy muon tracks from charged-current neutrino interactions at KM3NeT, we perform simulations using \texttt{TauRunner}~\cite{Safa:2021ghs}. This framework implements deep inelastic scattering at next-to-leading order through the CSMS model~\cite{Cooper-Sarkar:2011jtt}. The baseline CSMS cross section is known to omit corrections of $\mathcal{O}(10\%)$, which set the scale for the theoretical systematic uncertainties of our result. Specifically, NNLO QCD corrections and final-state radiation indicate that the CSMS cross section must be reduced by $\sim12\%$~\cite{Weigel:2024gzh}, whereas accounting for direct $W$-boson production can increase the total cross section by $\sim6\%$~\cite{Seckel:1997kk,Zhou:2019frk,Zhou:2019vxt}.

We adopt the Preliminary Reference Earth Model~\cite{Dziewonski:1981xy} for the Earth with a 4 km layer of water and place the KM3NeT detector at a depth of 3.4 km. 
Then, we inject an isotropic flux of muon neutrinos and simulate their scattering and muon energy losses. Since KM3NeT has only detected a single ultrahigh-energy event, our analysis remains valid even if it originated from a localized source. Because ultrahigh-energy transient sources are isotropically distributed, the probability of detecting a near-horizontal event is still dictated by Earth attenuation.

We take the initial energy of the neutrinos to follow a log-normal distribution centered around $220 \, \rm PeV$ and with a width of $\sigma \sim 0.43$. We have checked that a two-piece normal profile can improve our upper bound by roughly 20\%. Thus, a log-normal profile represents a conservative estimate of the unknown experimental likelihood profile, without qualitatively altering the physical conclusions. Following the analysis of the KM3NeT Collaboration, we discard any muons with energy less than 10 PeV. We take KM3NeT's geometric area and efficiency to be isotropic and independent of energy. 

In Fig.~\ref{fig:angular-distribution}, we show the simulated angular distribution of $>$10 PeV muon tracks in KM3NeT resulting from an isotropic flux of neutrinos, assuming the log-normal energy distribution centered at $220\, \rm PeV$.

These results are shown assuming only Standard Model interactions (filled green) and for total cross sections that are larger than the Standard Model prediction by factors of 10 and 50, respectively. Upgoing tracks in this energy range are strongly suppressed by neutrino absorption in the Earth, while the distribution is enhanced near the horizon due to the larger available target volume. 
Given this distribution (assuming only Standard Model interactions), it is unsurprising that the first neutrino-induced muon track in this energy range was observed near the horizon. If the total neutrino cross section were substantially larger, however, horizontal events would be suppressed, and a larger fraction of events would instead come from downgoing directions. 

\section{Constraints on the Total High-Energy Neutrino-Nucleon Cross Section}

Our analysis is based on an unbinned likelihood constructed from the predicted angular distribution,
\begin{equation}\label{eq:likelihood}
    \mathcal{L} = \int \dd \cos\theta\, \frac{\dd P_\mu}{\dd \cos\theta}\frac{1}{\sqrt{2\pi\Delta\theta}}\exp\left\{-\frac{(\theta-\theta_{\rm obs})^2}{2\Delta\theta^2}\right\},
\end{equation}
where ${\dd P_\mu}/{\dd \cos\theta}$ is the muon angular distribution shown in Fig.~\ref{fig:angular-distribution}, which is marginalized over neutrino energy. Then, $\theta_{\rm obs} = 90.6^\circ$ is the observed zenith angle of the KM3NeT event, and $\Delta\theta = 1.5^\circ$ its uncertainty. This likelihood quantifies the overlap between the arrival direction of the observed event and the predicted angular distribution.

\begin{figure}
    \centering
    \includegraphics[width=1\linewidth]{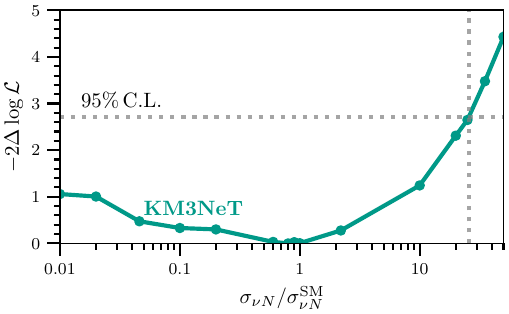}
    \caption{The change in the log-likelihood as a function of the total neutrino-nucleon cross section at $E_{\nu} = 220 \, {\rm PeV}$. 
    The dots represent points computed in the Markov Chain Monte Carlo chain and the solid line interpolates between them. \emph{The likelihood is maximized near the Standard Model value of the total cross section and the KM3NeT event allows us to constrain $\sigma_{\nu N} < 26 \, \sigma^{\rm SM}_{\nu N}$ at the 95\% confidence level.}}
    \label{fig:likelihood}
\end{figure}

In Fig.~\ref{fig:likelihood}, we plot the change in the log-likelihood as a function of the total neutrino-nucleon cross section, as evaluated at $E_{\nu}=220 \, {\rm PeV}$ and scaled to the value predicted by the Standard Model. At low cross sections, the Earth becomes transparent and the angular distribution becomes isotropic. As a result, the likelihood saturates at a constant baseline. This offset exists because a near-horizontal event is much better explained by a distribution that peaks near the horizon, rather than by an isotropic flux.
At the 68\% confidence level, the KM3NeT event yields a constraint of $\sigma_{\nu N}/\sigma^{\rm SM}_{\nu N} = 0.8^{+5.8}_{-0.78}$, entirely consistent with the Standard Model prediction. At the 95\% confidence level, this analysis yields a one-sided upper limit of $\sigma_{\nu N}/\sigma^{\rm SM}_{\nu N}<26$. If the cross section had exceeded this limit, a large fraction of any ultrahigh-energy neutrino-induced muon tracks would be downgoing, in tension with the near-horizontal direction of the KM3NeT event (see Fig.~\ref{fig:angular-distribution}).

A wide range of exotic physics scenarios have been proposed that would increase the total neutrino-nucleon cross section and consequently alter the predicted angular distribution of ultrahigh-energy muon tracks in large volume neutrino telescopes. That being said, we are not aware of any well-motivated physics scenarios that are not already ruled out by the Large Hadron Collider that would lead to cross sections as large as $\sigma_{\nu N}\gsim 26 \, \sigma^{\rm SM}_{\nu N}$.
As neutrino telescopes observe more events in this energy range, however, measurements of the neutrino-nucleon cross section are expected to become significantly more precise, potentially leading to constraints that will be competitive with those derived from accelerator-based experiments.

With this in mind, it is interesting to consider the projected sensitivity of future very large volume neutrino telescopes (for previous studies in this direction, see Refs.~\cite{Valera:2022ylt,Esteban:2022uuw}). Proposals for such telescopes include IceCube-Gen2~\cite{IceCube-Gen2:2020qha}, P-ONE~\cite{P-ONE:2020ljt}, TRIDENT~\cite{TRIDENT:2022hql}, HUNT~\cite{Huang:2023mzt}, GRAND 200k~\cite{GRAND:2018iaj}, BEACON 1k~\cite{Wissel:2020sec}, TAMBO~\cite{Romero-Wolf:2020pzh}, Trinity 18~\cite{Brown:2021tf}, RET-N~\cite{Prohira:2019glh} and POEMMA30~\cite{POEMMA:2020ykm}. The next generation of high- and ultrahigh-energy neutrino telescopes is expected to exceed the sensitivities of IceCube and KM3NeT by $\sim 1-2$ orders of magnitude across a wide range of energies~\cite{Ackermann:2022rqc}. 

In light of the upper limits placed by IceCube on the flux of ultrahigh-energy neutrinos, it seems likely that the 220 PeV event observed by KM3NeT represents a relatively unlikely upward fluctuation~\cite{KM3NeT:2025ccp,Li:2025tqf,Fang:2025nzg}. That being said, a future telescope such as IceCube-Gen2~\cite{IceCube-Gen2:2020qha} could be reasonably expected to detect many events in this energy range over the duration of its operation. Those events could be used to sample the angular distribution shown in Fig.~\ref{fig:angular-distribution}, and to improve the measurement of the total neutrino-nucleon cross section at energies beyond the reach of the Large Hadron Collider. 

\begin{figure}
    \centering
    \includegraphics[width=1\linewidth]{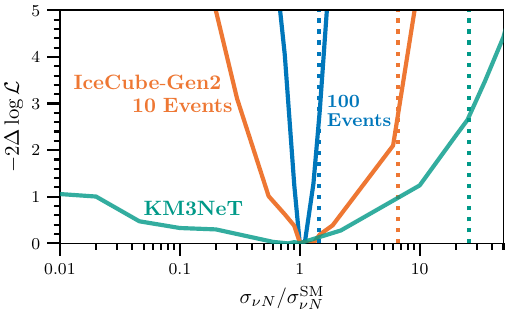}
    \caption{The change in the log-likelihood as a function of the total neutrino-nucleon cross section at $E_{\nu} =220 \, {\rm PeV}$, as projected from the observation of 10 or 100 neutrino-induced muon track events with IceCube-Gen2. A measurement of 10 such events is projected to result in a constraint of $\sigma_{\nu N}/\sigma_{\nu N}^{\rm SM} \lsim  6.5$ at the 95\% confidence level. With 100 events, this constraint is projected to improve to $\sigma_{\nu N}/\sigma_{\nu N}^{\rm SM} \lesssim 1.5$.}
    \label{fig:likelihood-forecast}
\end{figure}

To forecast the sensitivity of IceCube-Gen2 to the total neutrino-nucleon cross section, we assume a true value $\sigma_{\nu N} = \sigma_{\nu N}^{\rm SM}$ and carry out a Monte Carlo to simulate the angular distribution of observed muon tracks. In each trial, we assume that either 10 or 100 such events are observed, each arising from an isotropic flux of 220 PeV neutrinos. Note that the shallower depth of IceCube (which we take to be 2.0 km) leads to a somewhat smaller fraction of downgoing events than is shown in Fig.~\ref{fig:angular-distribution}. For IceCube-Gen2, we adopt an angular resolution of $\Delta\theta = 0.5^\circ$.

In Fig.~\ref{fig:likelihood-forecast}, we show the median projected constraint after observing 10 or 100 events and compare this to the upper limit we derived from the KM3NeT event. We find that IceCube-Gen2 could achieve a constraint of $\sigma_{\nu N}/\sigma_{\nu N}^{\rm SM} \lsim 6.5$ with 10 events at the 95\% confidence level, and $0.56\lesssim\sigma_{\nu N}/\sigma_{\nu N}^{\rm SM}\lesssim2.8$ at 68\% confidence. If IceCube-Gen2 observes 100 such events, an even stronger constraint of $\sigma_{\nu N}/\sigma_{\nu N}^{\rm SM} \lsim 1.5$ (95\%) could be obtained, corresponding to $0.92\lesssim\sigma_{\nu N}/\sigma_{\nu N}^{\rm SM}\lesssim1.25$ at 68\% confidence. 
Applying a similar analysis framework and set of assumptions, existing literature point out that such exposures can probe, for instance, scalar leptoquarks down to $M_{\rm LQ} \gtrsim 500\, \rm GeV$~\cite{Esteban:2022uuw} and leptogluons down to $M_{\rm LG} \gtrsim 3\, \rm TeV$~\cite{Bai:2025pef}.

\section{Summary and Conclusions}

In this work, we have used the ultrahigh-energy muon track event recently reported by the KM3NeT Collaboration to constrain the neutrino-nucleon cross section at center-of-momentum energies beyond the reach of terrestrial accelerators. The exceptional energy ($E_\nu \sim 220$ PeV) and near-horizontal direction of this event make it a powerful probe of exotic neutrino interactions. In particular, the predicted angular distribution of such events is sensitive to this cross section due to the absorption of neutrinos in the Earth.

By modeling the expected angular distribution of muon tracks induced by ultrahigh-energy neutrinos, we have constructed a likelihood function that quantifies the consistency of the observed event with different values of the total neutrino-nucleon cross section. From this analysis, we find that the data are fully consistent with the Standard Model prediction. At the 68\% confidence level, we obtain $\sigma_{\nu N}/\sigma_{\nu N}^{\rm SM} = 0.8^{+5.8}_{-0.78}$, while at the 95\% confidence level we derive a one-sided upper limit of $\sigma_{\nu N}/\sigma_{\nu N}^{\rm SM} < 26$ at $E_{\rm CM} \sim 20$ TeV. This represents the first constraint on the neutrino-nucleon cross section at center-of-momentum energies exceeding those directly probed by the Large Hadron Collider.

Although the constraint presented here is not yet competitive with accelerator-based bounds for well-motivated new physics scenarios, such measurements have the potential to provide a novel and independent probe of physics beyond the Standard Model. In particular, scenarios that predict large enhancements to the neutrino--nucleon cross section -- such as models with large or warped extra dimensions, leptoquarks, leptogluons, new gauge interactions, or non-perturbative electroweak effects -- can be constrained by the observed angular distribution of events induced by ultrahigh-energy neutrinos.

Looking ahead, the prospects for improving these measurements are highly promising. Next-generation neutrino observatories, such as IceCube-Gen2 and other proposed large-volume detectors, are expected to detect a significant number of events in this energy range. With even a modest sample of $\sim 10$ ultrahigh-energy muon tracks, we project that the neutrino-nucleon cross section could be constrained at a level of $\sigma_{\nu N}/\sigma_{\nu N}^{\rm SM} \lesssim 6.5$. This constraint could improve to $\sigma_{\nu N}/\sigma_{\nu N}^{\rm SM} \lesssim 1.5$ after $\sim 100$ such events. Such a measurement could be used to test a variety of scenarios beyond the Standard Model, and would be highly complementary to the information obtained from accelerator-based experiments.

\bigskip
\bigskip

\textit{Acknowledgements}--- We thank Caleb Gemmell and Josh Foster for useful discussions. This work has been supported by the Office of the Vice Chancellor for Research at the University of Wisconsin-Madison, with funding from the Wisconsin Alumni Research Foundation. 

\bibliography{bibliography}

@article{Safa:2021ghs,
    author = {Safa, Ibrahim and Lazar, Jeffrey and Pizzuto, Alex and Vasquez, Oswaldo and Arg{\"u}elles, Carlos A. and Vandenbroucke, Justin},
    title = "{TauRunner: A public Python program to propagate neutral and charged leptons}",
    eprint = "2110.14662",
    archivePrefix = "arXiv",
    primaryClass = "hep-ph",
    doi = "10.1016/j.cpc.2022.108422",
    journal = "Comput. Phys. Commun.",
    volume = "278",
    pages = "108422",
    year = "2022"
}

@article{Dziewonski:1981xy,
    author = "Dziewonski, A. M. and Anderson, D. L.",
    title = "{Preliminary reference earth model}",
    doi = "10.1016/0031-9201(81)90046-7",
    journal = "Phys. Earth Planet. Interiors",
    volume = "25",
    pages = "297--356",
    year = "1981"
}

@article{Ackermann:2022rqc,
    author = "Ackermann, Markus and others",
    title = "{High-energy and ultra-high-energy neutrinos: A Snowmass white paper}",
    eprint = "2203.08096",
    archivePrefix = "arXiv",
    primaryClass = "hep-ph",
    doi = "10.1016/j.jheap.2022.08.001",
    journal = "JHEAp",
    volume = "36",
    pages = "55--110",
    year = "2022"
}

@article{Prohira:2019glh,
    author = "Prohira, S. and others",
    title = "{Observation of Radar Echoes From High-Energy Particle Cascades}",
    eprint = "1910.12830",
    archivePrefix = "arXiv",
    primaryClass = "astro-ph.HE",
    doi = "10.1103/PhysRevLett.124.091101",
    journal = "Phys. Rev. Lett.",
    volume = "124",
    number = "9",
    pages = "091101",
    year = "2020"
}

@article{POEMMA:2020ykm,
    author = "Olinto, A. V. and others",
    collaboration = "POEMMA",
    title = "{The POEMMA (Probe of Extreme Multi-Messenger Astrophysics) observatory}",
    eprint = "2012.07945",
    archivePrefix = "arXiv",
    primaryClass = "astro-ph.IM",
    doi = "10.1088/1475-7516/2021/06/007",
    journal = "JCAP",
    volume = "06",
    pages = "007",
    year = "2021"
}

@article{GRAND:2018iaj,
    author = "{\'A}lvarez-Mu{\~n}iz, Jaime and others",
    collaboration = "GRAND",
    title = "{The Giant Radio Array for Neutrino Detection (GRAND): Science and Design}",
    eprint = "1810.09994",
    archivePrefix = "arXiv",
    primaryClass = "astro-ph.HE",
    doi = "10.1007/s11433-018-9385-7",
    journal = "Sci. China Phys. Mech. Astron.",
    volume = "63",
    number = "1",
    pages = "219501",
    year = "2020"
}

@article{Wissel:2020sec,
    author = "Wissel, Stephanie and others",
    title = "{Prospects for high-elevation radio detection of {\ensuremath{>}}100 PeV tau neutrinos}",
    eprint = "2004.12718",
    archivePrefix = "arXiv",
    primaryClass = "astro-ph.IM",
    doi = "10.1088/1475-7516/2020/11/065",
    journal = "JCAP",
    volume = "11",
    pages = "065",
    year = "2020"
}

@inproceedings{Romero-Wolf:2020pzh,
    author = "Romero-Wolf, Andres and others",
    title = "{An Andean Deep-Valley Detector for High-Energy Tau Neutrinos}",
    booktitle = "{Latin American Strategy Forum for Research Infrastructure}",
    eprint = "2002.06475",
    archivePrefix = "arXiv",
    primaryClass = "astro-ph.IM",
    month = "2",
    year = "2020"
}

@article{Brown:2021tf,
  author = "Brown, Anthony  and  Bagheri, Mahdi  and  Doro, Michele  and  Gazda, Eliza  and  Kieda, Dave  and  Lin, Chaoxian  and  Otte, Nepomuk  and  Taboada, Ignacio  and  Wang, Andrew",
  title = "{Trinity: an imaging air Cherenkov telescope to search for Ultra-High-Energy neutrinos.}",
  doi = "10.22323/1.395.1179",
  journal = "PoS",
  year = 2021,
  volume = "ICRC2021",
  pages = "1179"
}

@article{IceCube-Gen2:2020qha,
    author = "Aartsen, M. G. and others",
    collaboration = "IceCube-Gen2",
    title = "{IceCube-Gen2: the window to the extreme Universe}",
    eprint = "2008.04323",
    archivePrefix = "arXiv",
    primaryClass = "astro-ph.HE",
    doi = "10.1088/1361-6471/abbd48",
    journal = "J. Phys. G",
    volume = "48",
    number = "6",
    pages = "060501",
    year = "2021"
}

@article{P-ONE:2020ljt,
    author = "Agostini, Matteo and others",
    collaboration = "P-ONE",
    title = "{The Pacific Ocean Neutrino Experiment}",
    eprint = "2005.09493",
    archivePrefix = "arXiv",
    primaryClass = "astro-ph.HE",
    doi = "10.1038/s41550-020-1182-4",
    journal = "Nature Astron.",
    volume = "4",
    number = "10",
    pages = "913--915",
    year = "2020"
}

@article{TRIDENT:2022hql,
    author = "Ye, Z. P. and others",
    collaboration = "TRIDENT",
    title = "{A multi-cubic-kilometre neutrino telescope in the western Pacific Ocean}",
    eprint = "2207.04519",
    archivePrefix = "arXiv",
    primaryClass = "astro-ph.HE",
    doi = "10.1038/s41550-023-02087-6",
    journal = "Nature Astron.",
    volume = "7",
    number = "12",
    pages = "1497--1505",
    year = "2023"
}

@article{Huang:2023mzt,
    author = "Huang, Tian-Qi and Cao, Zhen and Chen, Mingjun and Liu, Jiali and Wang, Zike and You, Xiaohao and Qi, Ying",
    title = "{Proposal for the High Energy Neutrino Telescope}",
    doi = "10.22323/1.444.1080",
    journal = "PoS",
    volume = "ICRC2023",
    pages = "1080",
    year = "2023"
}

@article{Dutta:2000hh,
    author = "Dutta, S. Iyer and Reno, M. H. and Sarcevic, I. and Seckel, D.",
    title = "{Propagation of muons and taus at high-energies}",
    eprint = "hep-ph/0012350",
    archivePrefix = "arXiv",
    doi = "10.1103/PhysRevD.63.094020",
    journal = "Phys. Rev. D",
    volume = "63",
    pages = "094020",
    year = "2001"
}

@article{Gandhi:1995tf,
    author = "Gandhi, Raj and Quigg, Chris and Reno, Mary Hall and Sarcevic, Ina",
    title = "{Ultrahigh-energy neutrino interactions}",
    eprint = "hep-ph/9512364",
    archivePrefix = "arXiv",
    reportNumber = "FERMILAB-PUB-95-221-T, CLNS-95-1357, MRI-PHY-16-95, UIOWA-95-06, AZPH-TH-95-15",
    doi = "10.1016/0927-6505(96)00008-4",
    journal = "Astropart. Phys.",
    volume = "5",
    pages = "81--110",
    year = "1996"
}

@article{Pandey:2019apj,
    author = "Pandey, Sujata and Karmakar, Siddhartha and Rakshit, Subhendu",
    title = "{Strong constraints on non-standard neutrino interactions: LHC vs. IceCube}",
    eprint = "1907.07700",
    archivePrefix = "arXiv",
    primaryClass = "hep-ph",
    doi = "10.1007/JHEP11(2019)046",
    journal = "JHEP",
    volume = "11",
    pages = "046",
    year = "2019"
}

@article{Carena:1998gd,
    author = "Carena, Marcela and Choudhury, Debajyoti and Lola, Smaragda and Quigg, Chris",
    title = "{Manifestations of R-parity violation in ultrahigh-energy neutrino interactions}",
    eprint = "hep-ph/9804380",
    archivePrefix = "arXiv",
    reportNumber = "FERMILAB-PUB-98-088-T, MRI-PHY-980341, CERN-TH-98-97",
    doi = "10.1103/PhysRevD.58.095003",
    journal = "Phys. Rev. D",
    volume = "58",
    pages = "095003",
    year = "1998"
}

@article{Cornet:2001gy,
    author = "Cornet, F. and Illana, Jose I. and Masip, M.",
    title = "{TeV strings and the neutrino nucleon cross-section at ultrahigh-energies}",
    eprint = "hep-ph/0102065",
    archivePrefix = "arXiv",
    reportNumber = "UG-FT-127-01",
    doi = "10.1103/PhysRevLett.86.4235",
    journal = "Phys. Rev. Lett.",
    volume = "86",
    pages = "4235--4238",
    year = "2001"
}

@article{Anchordoqui:2006wc,
    author = "Anchordoqui, Luis A. and Garcia Canal, Carlos A. and Goldberg, Haim and Gomez Dumm, Daniel and Halzen, Francis",
    title = "{Probing leptoquark production at IceCube}",
    eprint = "hep-ph/0609214",
    archivePrefix = "arXiv",
    doi = "10.1103/PhysRevD.74.125021",
    journal = "Phys. Rev. D",
    volume = "74",
    pages = "125021",
    year = "2006"
}

@article{Becirevic:2018uab,
    author = "Be{\v{c}}irevi{\'c}, Damir and Panes, Boris and Sumensari, Olcyr and Zukanovich Funchal, Renata",
    title = "{Seeking leptoquarks in IceCube}",
    eprint = "1803.10112",
    archivePrefix = "arXiv",
    primaryClass = "hep-ph",
    doi = "10.1007/JHEP06(2018)032",
    journal = "JHEP",
    volume = "06",
    pages = "032",
    year = "2018"
}

@article{Chauhan:2017ndd,
    author = "Chauhan, Bhavesh and Kindra, Bharti and Narang, Ashish",
    title = "{Discrepancies in simultaneous explanation of flavor anomalies and IceCube PeV events using leptoquarks}",
    eprint = "1706.04598",
    archivePrefix = "arXiv",
    primaryClass = "hep-ph",
    doi = "10.1103/PhysRevD.97.095007",
    journal = "Phys. Rev. D",
    volume = "97",
    number = "9",
    pages = "095007",
    year = "2018"
}

@article{Mileo:2016zeo,
    author = "Mileo, Nicolas and de la Puente, Alejandro and Szynkman, Alejandro",
    title = "{Implications of a Electroweak Triplet Scalar Leptoquark on the Ultra-High Energy Neutrino Events at IceCube}",
    eprint = "1608.02529",
    archivePrefix = "arXiv",
    primaryClass = "hep-ph",
    doi = "10.1007/JHEP11(2016)124",
    journal = "JHEP",
    volume = "11",
    pages = "124",
    year = "2016"
}

@article{Dey:2016sht,
    author = "Dey, Ujjal Kumar and Mohanty, Subhendra and Tomar, Gaurav",
    title = "{Leptoquarks: 750 GeV Diphoton Resonance and IceCube Events}",
    eprint = "1606.07903",
    archivePrefix = "arXiv",
    primaryClass = "hep-ph",
    journal = "arXiv preprint",
    month = "6",
    year = "2016"
}

@article{Kirk:2023fin,
    author = "Kirk, Matthew and Okawa, Shohei and Wu, Keyun",
    title = "{A {\ensuremath{\nu}} window onto leptoquarks?}",
    eprint = "2307.11152",
    archivePrefix = "arXiv",
    primaryClass = "hep-ph",
    doi = "10.1007/JHEP12(2023)093",
    journal = "JHEP",
    volume = "12",
    pages = "093",
    year = "2023"
}

@article{Dey:2015eaa,
    author = "Dey, Ujjal Kumar and Mohanty, Subhendra",
    title = "{Constraints on Leptoquark Models from IceCube Data}",
    eprint = "1505.01037",
    archivePrefix = "arXiv",
    primaryClass = "hep-ph",
    doi = "10.1007/JHEP04(2016)187",
    journal = "JHEP",
    volume = "04",
    pages = "187",
    year = "2016"
}

@article{Denton:2020jft,
    author = "Denton, Peter B. and Kini, Yves",
    title = "{Ultra-High-Energy Tau Neutrino Cross Sections with GRAND and POEMMA}",
    eprint = "2007.10334",
    archivePrefix = "arXiv",
    primaryClass = "astro-ph.HE",
    doi = "10.1103/PhysRevD.102.123019",
    journal = "Phys. Rev. D",
    volume = "102",
    pages = "123019",
    year = "2020"
}

@article{Huang:2021mki,
    author = "Huang, Guo-yuan and Jana, Sudip and Lindner, Manfred and Rodejohann, Werner",
    title = "{Probing new physics at future tau neutrino telescopes}",
    eprint = "2112.09476",
    archivePrefix = "arXiv",
    primaryClass = "hep-ph",
    doi = "10.1088/1475-7516/2022/02/038",
    journal = "JCAP",
    volume = "02",
    number = "02",
    pages = "038",
    year = "2022"
}

@article{Zhou:2019frk,
    author = "Zhou, Bei and Beacom, John F.",
    title = "{W-boson and trident production in TeV{\textendash}PeV neutrino observatories}",
    eprint = "1910.10720",
    archivePrefix = "arXiv",
    primaryClass = "hep-ph",
    doi = "10.1103/PhysRevD.101.036010",
    journal = "Phys. Rev. D",
    volume = "101",
    number = "3",
    pages = "036010",
    year = "2020"
}

@article{Anchordoqui:2002vb,
    author = "Anchordoqui, Luis A. and Feng, Jonathan L. and Goldberg, Haim and Shapere, Alfred D.",
    title = "{Neutrino bounds on astrophysical sources and new physics}",
    eprint = "hep-ph/0207139",
    archivePrefix = "arXiv",
    reportNumber = "NUB-3229-TH-02, UCI-TR-2002-22, UK-02-09",
    doi = "10.1103/PhysRevD.66.103002",
    journal = "Phys. Rev. D",
    volume = "66",
    pages = "103002",
    year = "2002"
}

@article{Jain:2000pu,
    author = "Jain, P. and McKay, Douglas W. and Panda, S. and Ralston, John P.",
    title = "{Extra dimensions and strong neutrino nucleon interactions above 10**19-eV: Breaking the GZK barrier}",
    eprint = "hep-ph/0001031",
    archivePrefix = "arXiv",
    doi = "10.1016/S0370-2693(00)00647-X",
    journal = "Phys. Lett. B",
    volume = "484",
    pages = "267--274",
    year = "2000"
}

@article{Valera:2022ylt,
    author = "Valera, Victor Branco and Bustamante, Mauricio and Glaser, Christian",
    title = "{The ultra-high-energy neutrino-nucleon cross section: measurement forecasts for an era of cosmic EeV-neutrino discovery}",
    eprint = "2204.04237",
    archivePrefix = "arXiv",
    primaryClass = "hep-ph",
    doi = "10.1007/JHEP06(2022)105",
    journal = "JHEP",
    volume = "06",
    pages = "105",
    year = "2022"
}

@article{Morris:1993wg,
    author = "Morris, D. A. and Ringwald, A.",
    title = "{Cosmic ray signatures of multi - W processes}",
    eprint = "hep-ph/9308269",
    archivePrefix = "arXiv",
    reportNumber = "CERN-TH-6822-93, UCLA-93-TEP-24",
    doi = "10.1016/0927-6505(94)90017-5",
    journal = "Astropart. Phys.",
    volume = "2",
    pages = "43--66",
    year = "1994"
}

@article{Fodor:2003bn,
    author = "Fodor, Z. and Katz, S. D. and Ringwald, A. and Tu, H.",
    title = "{Electroweak instantons as a solution to the ultrahigh-energy cosmic ray puzzle}",
    eprint = "hep-ph/0303080",
    archivePrefix = "arXiv",
    reportNumber = "WUB-03-03, ITP-BUDAPEST-593, DESY-03-022",
    doi = "10.1016/S0370-2693(03)00487-8",
    journal = "Phys. Lett. B",
    volume = "561",
    pages = "191--201",
    year = "2003"
}

@article{Ringwald:2001vk,
    author = "Ringwald, A. and Tu, H.",
    title = "{Collider versus cosmic ray sensitivity to black hole production}",
    eprint = "hep-ph/0111042",
    archivePrefix = "arXiv",
    reportNumber = "DESY-01-182",
    doi = "10.1016/S0370-2693(01)01421-6",
    journal = "Phys. Lett. B",
    volume = "525",
    pages = "135--142",
    year = "2002"
}

@article{Illana:2004qc,
    author = "Illana, J. I. and Masip, M. and Meloni, D.",
    title = "{Cosmogenic neutrinos and signals of TeV gravity in air showers and neutrino telescopes}",
    eprint = "hep-ph/0402279",
    archivePrefix = "arXiv",
    reportNumber = "CAFPE-31-04, UG-FT-161-04",
    doi = "10.1103/PhysRevLett.93.151102",
    journal = "Phys. Rev. Lett.",
    volume = "93",
    pages = "151102",
    year = "2004"
}

@article{Dutta:2015dka,
    author = "Dutta, Bhaskar and Gao, Yu and Li, Tianjun and Rott, Carsten and Strigari, Louis E.",
    title = "{Leptoquark implication from the CMS and IceCube experiments}",
    eprint = "1505.00028",
    archivePrefix = "arXiv",
    primaryClass = "hep-ph",
    reportNumber = "MI-TH-1509",
    doi = "10.1103/PhysRevD.91.125015",
    journal = "Phys. Rev. D",
    volume = "91",
    pages = "125015",
    year = "2015"
}

@article{Ellis:2016dgb,
    author = "Ellis, John and Sakurai, Kazuki and Spannowsky, Michael",
    title = "{Search for Sphalerons: IceCube vs. LHC}",
    eprint = "1603.06573",
    archivePrefix = "arXiv",
    primaryClass = "hep-ph",
    reportNumber = "KCL-PH-TH-2016-13, LCTS-2016-08, CERN-TH-2016-064, IPPP-16-24",
    doi = "10.1007/JHEP05(2016)085",
    journal = "JHEP",
    volume = "05",
    pages = "085",
    year = "2016"
}

@article{Alvarez-Muniz:2001efi,
    author = "Alvarez-Muniz, J. and Halzen, F. and Han, Tao and Hooper, D.",
    title = "{Phenomenology of high-energy neutrinos in low scale quantum gravity models}",
    eprint = "hep-ph/0107057",
    archivePrefix = "arXiv",
    reportNumber = "MADPH-01-1236",
    doi = "10.1103/PhysRevLett.88.021301",
    journal = "Phys. Rev. Lett.",
    volume = "88",
    pages = "021301",
    year = "2002"
}

@article{Alvarez-Muniz:2002snq,
    author = "Alvarez-Muniz, Jaime and Feng, Jonathan L. and Halzen, Francis and Han, Tao and Hooper, Dan",
    title = "{Detecting microscopic black holes with neutrino telescopes}",
    eprint = "hep-ph/0202081",
    archivePrefix = "arXiv",
    reportNumber = "MIT-CTP-3221, UCI-TR-2001-43, MADPH-02-1255",
    doi = "10.1103/PhysRevD.65.124015",
    journal = "Phys. Rev. D",
    volume = "65",
    pages = "124015",
    year = "2002"
}

@article{Jain:2002kz,
    author = "Jain, Pankaj and Kar, Supriya and McKay, Douglas W. and Panda, Sukanta and Ralston, John P.",
    title = "{Angular dependence of neutrino flux in KM**3 detectors in low scale gravity models}",
    eprint = "hep-ph/0205052",
    archivePrefix = "arXiv",
    doi = "10.1103/PhysRevD.66.065018",
    journal = "Phys. Rev. D",
    volume = "66",
    pages = "065018",
    year = "2002"
}

@article{Anchordoqui:2003jr,
    author = "Anchordoqui, Luis A. and Feng, Jonathan L. and Goldberg, Haim and Shapere, Alfred D.",
    title = "{Updated limits on TeV scale gravity from absence of neutrino cosmic ray showers mediated by black holes}",
    eprint = "hep-ph/0307228",
    archivePrefix = "arXiv",
    reportNumber = "NUB-3239-TH-03, UCI-TR-2003-31, UK-03-10",
    doi = "10.1103/PhysRevD.68.104025",
    journal = "Phys. Rev. D",
    volume = "68",
    pages = "104025",
    year = "2003"
}

@article{Friess:2002cc,
    author = "Friess, Joshua J. and Han, Tao and Hooper, Dan",
    title = "{TeV string state excitation via high-energy cosmic neutrinos}",
    eprint = "hep-ph/0204112",
    archivePrefix = "arXiv",
    doi = "10.1016/S0370-2693(02)02728-4",
    journal = "Phys. Lett. B",
    volume = "547",
    pages = "31--36",
    year = "2002"
}

@article{Han:2003ru,
    author = "Han, Tao and Hooper, Dan",
    title = "{Effects of electroweak instantons in high-energy neutrino telescopes}",
    eprint = "hep-ph/0307120",
    archivePrefix = "arXiv",
    reportNumber = "MADPH-03-1338",
    doi = "10.1016/j.physletb.2003.12.040",
    journal = "Phys. Lett. B",
    volume = "582",
    pages = "21--26",
    year = "2004"
}

@article{Han:2004kq,
    author = "Han, Tao and Hooper, Dan",
    title = "{The Particle physics reach of high-energy neutrino astronomy}",
    eprint = "hep-ph/0408348",
    archivePrefix = "arXiv",
    reportNumber = "MADPH-04-1392",
    doi = "10.1088/1367-2630/6/1/150",
    journal = "New J. Phys.",
    volume = "6",
    pages = "150",
    year = "2004"
}

@article{IceCube:2020rnc,
    author = "Abbasi, R. and others",
    collaboration = "IceCube",
    title = "{Measurement of the high-energy all-flavor neutrino-nucleon
cross section with IceCube}",
    eprint = "2011.03560",
    archivePrefix = "arXiv",
    primaryClass = "hep-ex",
    doi = "10.1103/PhysRevD.104.022001",
    journal = "Phys. Rev. D",
    volume = "104",
    pages = "022001",
    year = "2021"
}

@article{Anchordoqui:2005pn,
    author = "Anchordoqui, Luis A. and Feng, Jonathan L. and Goldberg, Haim",
    title = "{Particle physics on ice: Constraints on neutrino interactions far above the weak scale}",
    eprint = "hep-ph/0504228",
    archivePrefix = "arXiv",
    reportNumber = "NUB-3255-TH-05, UCI-TR-2005-13",
    doi = "10.1103/PhysRevLett.96.021101",
    journal = "Phys. Rev. Lett.",
    volume = "96",
    pages = "021101",
    year = "2006"
}

@article{IceCube:2017roe,
    author = "Aartsen, M. G. and others",
    collaboration = "IceCube",
    title = "{Measurement of the multi-TeV neutrino cross section with IceCube using Earth absorption}",
    eprint = "1711.08119",
    archivePrefix = "arXiv",
    primaryClass = "hep-ex",
    doi = "10.1038/nature24459",
    journal = "Nature",
    volume = "551",
    pages = "596--600",
    year = "2017"
}

@article{Borriello:2007cs,
    author = "Borriello, E. and Cuoco, A. and Mangano, G. and Miele, G. and Pastor, S. and Pisanti, O. and Serpico, P. D.",
    title = "{Disentangling neutrino-nucleon cross section and high energy neutrino flux with a km$^3$ neutrino telescope}",
    eprint = "0711.0152",
    archivePrefix = "arXiv",
    primaryClass = "astro-ph",
    reportNumber = "DSF-36-2007, FERMILAB-PUB-07-582-A, IFIC-07-60",
    doi = "10.1103/PhysRevD.77.045019",
    journal = "Phys. Rev. D",
    volume = "77",
    pages = "045019",
    year = "2008"
}

@inbook{Klein:2019nbu,
    author = "Klein, Spencer R.",
    title = "{Probing high-energy interactions of atmospheric and astrophysical neutrinos}",
    eprint = "1906.02221",
    archivePrefix = "arXiv",
    primaryClass = "astro-ph.HE",
    doi = "10.1142/9789813275027_0004",
    year = "2020"
}

@article{Bustamante:2017xuy,
    author = "Bustamante, Mauricio and Connolly, Amy",
    title = "{Extracting the Energy-Dependent Neutrino-Nucleon Cross Section above 10 TeV Using IceCube Showers}",
    eprint = "1711.11043",
    archivePrefix = "arXiv",
    primaryClass = "astro-ph.HE",
    doi = "10.1103/PhysRevLett.122.041101",
    journal = "Phys. Rev. Lett.",
    volume = "122",
    number = "4",
    pages = "041101",
    year = "2019"
}

@article{Kusenko:2001gj,
    author = "Kusenko, Alexander and Weiler, Thomas J.",
    title = "{Neutrino cross-sections at high-energies and the future observations of ultrahigh-energy cosmic rays}",
    eprint = "hep-ph/0106071",
    archivePrefix = "arXiv",
    reportNumber = "UCLA-01-TEP-11, VAND-TH-01-06",
    doi = "10.1103/PhysRevLett.88.161101",
    journal = "Phys. Rev. Lett.",
    volume = "88",
    pages = "161101",
    year = "2002"
}

@article{Hooper:2002yq,
    author = "Hooper, Dan",
    title = "{Measuring high-energy neutrino nucleon cross-sections with future neutrino telescopes}",
    eprint = "hep-ph/0203239",
    archivePrefix = "arXiv",
    reportNumber = "MADPH-02-1261",
    doi = "10.1103/PhysRevD.65.097303",
    journal = "Phys. Rev. D",
    volume = "65",
    pages = "097303",
    year = "2002"
}

@article{KM3NeT:2025npi,
    author = "Aiello, S. and others",
    collaboration = "KM3NeT",
    title = "{Observation of an ultra-high-energy cosmic neutrino with KM3NeT}",
    doi = "10.1038/s41586-024-08543-1",
    journal = "Nature",
    volume = "638",
    number = "8050",
    pages = "376--382",
    year = "2025",
    note = "[Erratum: Nature 640, E3 (2025)]"
}

@article{KM3NeT:2025aps,
    author = "Adriani, O. and others",
    collaboration = "KM3NeT",
    title = "{On the Potential Galactic Origin of the Ultra-High-Energy Event KM3-230213A}",
    eprint = "2502.08387",
    archivePrefix = "arXiv",
    primaryClass = "astro-ph.HE",
    month = "2",
    journal = "arXiv preprint",
    year = "2025"
}

@article{KM3NeT:2025ccp,
    author = "Adriani, O. and others",
    collaboration = "KM3NeT",
    title = "{Ultrahigh-Energy Event KM3-230213A within the Global Neutrino Landscape}",
    eprint = "2502.08173",
    archivePrefix = "arXiv",
    primaryClass = "astro-ph.HE",
    doi = "10.1103/yypk-zmb8",
    journal = "Phys. Rev. X",
    volume = "15",
    number = "3",
    pages = "031016",
    year = "2025"
}

\end{document}